\documentclass[american,aps,pra,showpacs,preprint]{revtex4}
\usepackage{times}
\usepackage[T1]{fontenc}
\usepackage[latin1]{inputenc}
\usepackage{color}
\usepackage{graphicx}
\usepackage{amssymb}

\makeatletter

\renewcommand{\vec}{\mathbf}

\usepackage{babel}
\makeatother
\begin{document}

\title{Quantum recoil effects in finite-time disentanglement of two distinguishable
atoms}

\author{F. Lastra, S. Wallentowitz, M. Orszag, and M. Hernández}

\affiliation{Facultad de Física, Pontificia Universidad Católica de Chile, Casilla
306, Santiago 22, Chile}


\begin{abstract}
Starting from the requirement of distinguishability of two atoms by
their positions, it is shown that photon recoil has a strong influence
on finite-time disentanglement and in some cases prevents its appearance.
At near-field inter atomic distances well localized atoms --- with
maximally one atom being initially excited --- may suffer disentanglement
at a single finite time or even at a series of equidistant finite
times, depending on their mean inter atomic distance and their initial
electronic preparation. 
\end{abstract}
\maketitle

\section{Introduction}

As the interaction of a quantum system with its environment may be
suppressed but never completely removed, the system's coherence will
always suffer a degrading, i.e. decoherence \cite{joos}. Regarding
the system as being composed of two distinguishable constituents,
apart from the system's coherence the entanglement \cite{schroedi}
between constituents also suffers a degrading. However, whereas coherence
is gradually lost and asymptotically decays to zero at infinite time,
entanglement may disappear even at \emph{finite} time and in an abrupt
way \cite{esd}.

In recent years, there has been an increased interest in finite-time
disentanglement with bipartite systems \cite{esd2,non-markovian,tanas,eberly3,orszag},
a phenomenon that has also been experimentally observed \cite{almeida,almeida2}.
However, to best of our knowledge, no emphasis has been given to the
question how to actually distinguish between the constituents. Their
distinguishability is required to rightfully apply the concept of
entanglement between them. Distinguishability requires the use of
an additional degree of freedom that usually takes part in the system's
dynamics. Up to now, this somewhat hidden resource has been disregarded
in the literature. However, its inclusion is required for a complete
physical description. As is shown in this Letter, such a more complete
physical description will lead to strong modifications with respect
to the appearance of finite-time disentanglement.

The outline of the paper is as follows: In Sec. \ref{sec:Distinguishability-of-entangled}
the requirement of distinguishability of entangled atoms is developped,
which shows the need for treating correctly the relative quantum motion
of atoms and the photon recoil. The solution for the electronic probability
amplitudes of the atoms is then obtained in Sec. \ref{sec:Interaction-of-two},
assuming an initial single excitation in the atom-field system. Finally,
the finite-time disentanglement conditions are discussed in Sec. \ref{sec:Finite-Time-Disentanglement}
and a summary and conclusions are given in Sec. \ref{sec:Summary-and-conclusions}.

\section{Distinguishability of entangled atoms\label{sec:Distinguishability-of-entangled}}

Consider two identical two-level atoms that are supposed to be distinguishable
by their positions. Maintaining distinguishability during a duration
of the order of the natural lifetime $\tau_{0}$ of the electronic
excited state requires that the quantum dispersion of the relative-position
wave-packet be sufficiently weak. The wave-packet should be well localized
at all times, otherwise the atoms could no longer be distinguished
by their positions. More precisely, the condition for distinguishability
of the atoms is, that within the time duration $\sim\tau_{0}$, \begin{equation}
\Delta r\ll r,\label{eq:distinguishability}\end{equation}
 where $\Delta r$ and $r$ are rms spread and mean of the distance
between the atoms, respectively.

In the absence of relative motion, i.e. at a mean inter atomic distance
$r_{0}$, the initial rms spread $\Delta r_{0}$ is enlarged by quantum
dispersion during the excited-state lifetime $\tau_{0}$ to\begin{equation}
\Delta r=\Delta r_{0}\sqrt{1+\left(\frac{l}{\Delta r_{0}}\right)^{4}}.\label{eq:dispersion}\end{equation}
Here the dispersion length is defined as \begin{equation}
l=\sqrt{\hbar\tau_{0}/m},\end{equation}
where $m$ is the atomic mass and $\tau_{0}=2\pi/\gamma_{0}$ with
\begin{equation}
\gamma_{0}=\frac{d^{2}\omega_{0}^{3}}{6\pi\epsilon_{0}\hbar c^{3}}\end{equation}
being the natural line width of the atom's electronic transition with
dipole moment $\vec{d}$ and transition frequency $\omega_{0}$.

Using Eq. (\ref{eq:dispersion}), the condition (\ref{eq:distinguishability})
then leads to \begin{equation}
\Delta r_{0}^{4}-\Delta r_{0}^{2}r_{0}^{2}+l^{4}\ll0,\end{equation}
which has solutions only for mean interatomic distances $r_{0}\gg l$,
which then establish limits for the rms spread: \begin{equation}
\Delta r_{{\rm min}}\ll\Delta r_{0}\ll r_{0}.\label{eq:dr-limits}\end{equation}
The minimum rms spread is \begin{equation}
\Delta r_{{\rm min}}=l^{2}/r_{0}\ll l,\end{equation}
which can be written as \begin{equation}
\Delta r_{{\rm min}}\ll\lambda_{0}\sqrt{\frac{E_{{\rm r}}}{\hbar\gamma_{0}}}.\end{equation}

As the recoil energy $E_{{\rm r}}=(\hbar k_{0})^{2}/2m$ is typically
smaller than $\hbar\gamma_{0}$, the minimum spread may still be much
smaller than the transition wavelength $\lambda_{0}$ so that rather
small spreads are allowed. However, a zero spread is not permitted,
as then quantum dispersion would render the atoms indistinguishable.
Thus, to be consistent with the requirement of the atoms being distinguishable
during the excited-state lifetime $\tau_{0}$, a finite initial spread
$\Delta r_{0}$ within the limits (\ref{eq:dr-limits}) is required.

\section{Interaction of two atoms with the electromagnetic field\label{sec:Interaction-of-two}}

\subsection{Hamilton operator of the atom-field system}

The Hamiltonian describing the free radiation field, the two atoms
with their corresponding kinetic energies, and the atom-field interaction
is \begin{equation}
\hat{H}=\int d^{3}k\sum_{\sigma}\hbar ck\hat{n}_{\vec{k},\sigma}+\sum_{a=\pm}\hbar\omega_{0}\hat{S}_{a,z}+\frac{\hat{\vec{P}}^{2}}{4m}+\frac{\hat{\vec{p}}^{2}}{m}+\hat{V}.\label{eq:H0}\end{equation}
Here $\hat{n}_{\vec{k},\sigma}=\hat{a}_{\vec{k},\sigma}^{\dagger}\hat{a}_{\vec{k},\sigma}$
is the photon-number operator with $\hat{a}_{\vec{k},\sigma}$ being
the bosonic annihilation operator of a photon in mode $(\vec{k},\sigma)$,
$\sigma$ denoting one of the two polarizations orthogonal to the
wave vector $\vec{k}$. The electronic two-level systems with transition
frequency $\omega_{0}$ are described by the pseudo spin operators
$\hat{\vec{S}}_{a}$, where the index $a=\pm$ indicates the atom
under consideration, and $\hat{\vec{P}}$ and $\hat{\vec{p}}$ are
center-of-mass and relative momentum, respectively.

The interaction part $\hat{V}$ in Eq. (\ref{eq:H0}) describes the
photon absorption and emission processes associated with recoil on
the corresponding atom. It can be written as \begin{equation}
\hat{V}=\int d^{3}k\sum_{\sigma}\sum_{a=\pm}\hbar\kappa_{\vec{k},\sigma}\hat{S}_{a,+}\hat{a}_{\vec{k},\sigma}e^{i\vec{k}\cdot\hat{\vec{R}}-ia\vec{k}\cdot\hat{\vec{r}}/2}+{\rm H.a.},\end{equation}
 where $\hat{\vec{R}}$ and $\hat{\vec{r}}$ are the center-of-mass
and relative position of the atoms, respectively, and the vacuum Rabi
frequency of the electromagnetic mode $(\vec{k},\sigma)$ is \begin{equation}
\kappa_{\vec{k},\sigma}=\vec{d}\cdot\vec{e}_{\sigma}(\vec{k})E_{k}/\hbar,\end{equation}
with the polarization unit vector $\vec{e}_{\sigma}(\vec{k})$ and
the rms electric-field vacuum fluctuation \begin{equation}
E_{k}=\sqrt{\frac{\hbar ck}{16\pi^{3}\epsilon_{0}}}.\end{equation}

\subsection{Wigner--Weisskopf solution for an initial single excitation}

As we assume the electromagnetic field to be initially in its vacuum
state with only one of the atoms being excited, we take the general
form of the quantum state as:\begin{eqnarray}
 &  & |\Psi(t)\rangle=\int d^{3}p\int d^{3}P|\vec{p}\rangle_{{\rm rel}}\otimes|\vec{P}\rangle_{{\rm cm}}\label{eq:Psi-ansatz}\\
 &  & \otimes\left[\sum_{a=\pm}\psi_{a}(\vec{p},\vec{P},t)|a\rangle+\int d^{3}k\sum_{\sigma}\psi_{\vec{k},\sigma}(\vec{p},\vec{P},t)|\vec{k},\sigma\rangle\right].\nonumber \end{eqnarray}
 Here the state with atom $a=\pm$ being excited and no photon being
present and the state with no atom being excited but a photon in mode
$(\vec{k},\sigma)$ being present, are defined as\begin{eqnarray}
|\pm\rangle & = & |{\textstyle \mp\frac{1}{2}},{\textstyle \pm\frac{1}{2}}\rangle_{{\rm el}}\otimes|{\rm vac}\rangle_{{\rm em}},\\
|\vec{k},\sigma\rangle & = & |{\textstyle -\frac{1}{2}},{\textstyle -\frac{1}{2}}\rangle_{{\rm el}}\otimes|\vec{k},\sigma\rangle_{{\rm em}},\end{eqnarray}
 respectively, where \begin{equation}
|m,m^{\prime}\rangle_{{\rm el}}=|m\rangle_{-}\otimes|m^{\prime}\rangle_{+}.\end{equation}

From the Schrödinger equation the equations of motion for the probability
amplitudes in Eq. (\ref{eq:Psi-ansatz}) are easily derived and Laplace
transformed ($t\to s$). Eliminating then the photon probability amplitude
$\psi_{\vec{k},\sigma}$, the equations for the probability amplitudes
for one of the atoms being excited are obtained as:\begin{eqnarray}
 &  & \left[s+\Gamma(\vec{p}\pm\hbar\vec{k}/2,\vec{P}-\hbar\vec{k},s)+i\frac{p^{2}+P^{2}/4}{\hbar m}\right]\underline{\psi}_{\pm}(\vec{p},\vec{P},s)\nonumber \\
 &  & +\int d^{3}k\gamma(\vec{k};\vec{p}\pm\hbar\vec{k}/2,\vec{P}-\hbar\vec{k},s)\underline{\psi}_{\mp}(\vec{p}\pm\hbar\vec{k},\vec{P},s)\nonumber \\
 &  & =\psi_{\pm}(\vec{p},\vec{P},0),\label{eq:psia}\end{eqnarray}
where $\psi_{\pm}(\vec{p},\vec{P},0)$ is the initial probability
amplitude at time $t=0$. The complex-valued spectral rate is defined
as\begin{equation}
\gamma(\vec{k};\vec{p},\vec{P},s)=\frac{\sum_{\sigma}\kappa_{\vec{k},\sigma}\kappa_{\vec{k},\sigma}^{\ast}}{s+i[ck-\omega_{0}+(p^{2}+P^{2}/4)/\hbar m]},\end{equation}
 with the integrated rate being \begin{equation}
\Gamma(\vec{p},\vec{P},s)=\int d^{3}k\gamma(\vec{k};\vec{p},\vec{P},s).\end{equation}

Given that relative and center-of-mass kinetic energies, as well as
the recoil energy $E_{{\rm r}}$, typically produce frequency shifts
much smaller than the natural line width $\gamma_{0}$,\begin{equation}
\frac{p^{2}}{m},\frac{P^{2}}{4m},E_{{\rm r}}\ll\hbar\gamma_{0},\label{eq:cond2}\end{equation}
the required rates in Eq. (\ref{eq:psia}) can be approximated as
being independent of the atoms momenta, \begin{equation}
\gamma(\vec{k};\vec{p}\pm\hbar\vec{k}/2,-\hbar\vec{k},s)\approx\gamma(\vec{k};s)=\frac{\sum_{\sigma}\kappa_{\vec{k},\sigma}\kappa_{\vec{k},\sigma}^{\ast}}{s+i(ck-\omega_{0})},\label{eq:gamma-approx}\end{equation}
 and correspondingly\begin{equation}
\Gamma(\vec{p}\pm\hbar\vec{k}/2,\vec{P}-\hbar\vec{k},s)\approx\Gamma(s)=\int d^{3}k\gamma(\vec{k};s).\label{eq:Gamma-approx}\end{equation}

Using the approximated rates (\ref{eq:gamma-approx}) and (\ref{eq:Gamma-approx}),
Eqs (\ref{eq:psia}) can be diagonalized by the use of the amplitudes\begin{equation}
\underline{\phi}_{\pm}(\vec{p},\vec{P},s)=\frac{\underline{\psi}_{+}(\vec{p},\vec{P},s)\pm\underline{\psi}_{-}(\vec{p},\vec{P},s)}{\sqrt{2}},\end{equation}
 and Fourier transformed ($\vec{p}\to\vec{r}$, $\vec{P}\to\vec{R}$)
to obtain\begin{eqnarray}
 &  & \left[s+\Gamma(s)\pm\gamma(\vec{r};s)-i\left(\frac{\hbar\nabla_{r}^{2}}{m}+\frac{\hbar\nabla_{R}^{2}}{4m}\right)\right]\underline{\phi}_{\pm}(\vec{r},\vec{R},s)\nonumber \\
 &  & \qquad=\phi_{\pm}(\vec{r},\vec{R},0).\label{eq:phi-eq}\end{eqnarray}

Non-Markovian effects \cite{non-markovian} may be safely neglected
by taking the limit $s\to0$ for the Fourier transform of the rate
(\ref{eq:gamma-approx}), which is obtained as \begin{equation}
\gamma(\vec{r})=\lim_{s\to0}\gamma(\vec{r};s)=\gamma_{0}\left[\mu(k_{0}\vec{r})+i\nu(k_{0}\vec{r})\right].\label{eq:gamma-r2}\end{equation}
 Here $\mu$ and $\nu$ are given by \begin{eqnarray}
\mu(\vec{x}) & = & \frac{3}{2}\left[\left(3\varsigma-2\right)\left(\frac{\cos x}{x^{2}}-\frac{\sin x}{x^{3}}\right)+\varsigma\frac{\sin x}{x}\right],\label{eq:mu-def}\\
\nu(\vec{x}) & = & \frac{3}{4}\left[\left(3\varsigma-2\right)\left(\frac{\sin x}{x^{2}}+\frac{\cos x}{x^{3}}\right)-\varsigma\frac{\cos x}{x}\right],\label{eq:nu-def}\end{eqnarray}
where $\varsigma=\sin^{2}\theta$ with $\theta$ being the angle between
$\vec{x}$ and the dipole transition moment $\vec{d}$. Whereas the
imaginary part of Eq. (\ref{eq:gamma-r2}) describes the dipole-dipole
interaction between the atoms, the real part is responsible for a
dependence of collective spontaneous emission of both atoms on their
distance and the dipole orientation \cite{lehmberg,lehmberg2}. As
$\Gamma=\gamma(\vec{0})$ and $\mu(\vec{x})\to1$ for $x\to0$, the
real part of the integrated rate becomes $\Re(\Gamma)=\gamma_{0}$.

According to Eq. (\ref{eq:cond2}), the kinetic energy terms in Eq.
(\ref{eq:phi-eq}) are much smaller than $\gamma_{0}$, the latter
appearing in the equation via $\gamma(\vec{r})$ and $\Gamma$. Thus
we consistently neglect those terms and obtain the solution of Eq.
(\ref{eq:phi-eq}), whose inverse Laplace transform then gives the
probability amplitude for atom $a=\pm$ being excited in terms of
the initial amplitudes:\begin{equation}
\psi_{\pm}(\vec{r},\vec{R},t)=\frac{\phi_{+}(\vec{r},\vec{R},0)e^{-\gamma(\vec{r})t}\pm\phi_{-}(\vec{r},\vec{R},0)e^{\gamma(\vec{r})t}}{\sqrt{2}}e^{-\gamma_{0}t}.\end{equation}

\section{Finite-time disentanglement\label{sec:Finite-Time-Disentanglement}}

\subsection{Concurrence in terms of the moments of $\mu$ and $\nu$ }

For obtaining the entanglement between the atoms electronic subsystems
we require the reduced electronic density operator,\[
\hat{\rho}_{{\rm el}}={\rm Tr}_{{\rm rel}}{\rm Tr}_{{\rm cm}}{\rm Tr}_{{\rm em}}\left[|\Psi(t)\rangle\langle\Psi(t)|\right],\]
 whose density matrix is\begin{equation}
\rho_{{\rm el}}(t)=\left(\begin{array}{cccc}
0 & 0 & 0 & 0\\
0 & p_{-}(t) & z(t) & 0\\
0 & z^{\ast}(t) & p_{+}(t) & 0\\
0 & 0 & 0 & p(t)\end{array}\right)\label{eq:matrix}\end{equation}
in the standard basis \[
\{|{\textstyle \frac{1}{2},{\textstyle \frac{1}{2}}\rangle},|{\textstyle \frac{1}{2},{\textstyle -\frac{1}{2}}\rangle},|{\textstyle -\frac{1}{2},{\textstyle \frac{1}{2}}\rangle},|{\textstyle -\frac{1}{2},{\textstyle -\frac{1}{2}}\rangle}\}.\]
Here $p(t)=1-p_{+}(t)-p_{-}(t)$ with $p_{a}(t)$ being the probability
for atom $a=\pm$ being excited and the non-diagonal element $z$
being a coherence/correlation between the atoms. Given that initially
only one excitation exists in the system, the first diagonal element
in (\ref{eq:matrix}) vanished and thus the concurrence \cite{wooters}
--- as a measure of entanglement --- is simply

\[
C(t)=2\max\{0,|z(t)|\},\]
 where the non-diagonal element is obtained as\begin{equation}
z(t)=\int d^{3}r\int d^{3}R\psi_{-}(\vec{r},\vec{R},t)\psi_{+}^{\ast}(\vec{r},\vec{R},t).\label{eq:coherence}\end{equation}

For the initial quantum state we choose a state, where relative and
center-of-mass motion factorize from the possibly non-factorisable
electronic state:\[
\psi_{a}(\vec{r},\vec{R},0)=\Psi_{a}\sqrt{p_{{\rm rel}}(\vec{r})p_{{\rm cm}}(\vec{R})},\]
 where $p_{{\rm rel}}$ and $p_{{\rm cm}}$ is the initial probability
density for the relative and center-of-mass coordinate, respectively.
Using this form, the matrix element (\ref{eq:coherence}) results
as\begin{eqnarray}
z(t) & = & \int d^{3}rp_{{\rm rel}}(\vec{r})\left\{ \sum_{a=\pm}\frac{a}{2}|\Phi_{a}|^{2}e^{-2\gamma_{0}t[1+a\mu(\vec{r})]}\right.\nonumber \\
 &  & \left.+i\Im\left[\Phi_{+}\Phi_{-}^{\ast}e^{-2\gamma_{0}t[1-i\nu(\vec{r})]}\right]\right\} ,\label{eq:z-def}\end{eqnarray}
 with the initial electronic amplitudes \[
\Phi_{\pm}=(\Psi_{+}\pm\Psi_{-})/\sqrt{2}.\]

For averaging over the inter atomic distance we choose for $p_{{\rm rel}}$
a normalized isotropic Gaussian with mean distance $\vec{r}_{0}$
and rms spread $\Delta r_{0}$, as introduced previously. Given that
the time of disentanglement is expected to be smaller than the excited-state
lifetime, i.e. $\gamma_{0}t\ll1$, the real and imaginary parts of
the coherence (\ref{eq:z-def}) can be obtained in second-order cumulant
expansion as\begin{eqnarray}
z_{r}(t) & = & \frac{1}{2}\left(|\Phi_{+}|^{2}e^{-2\bar{\mu}\gamma_{0}t}-|\Phi_{-}|^{2}e^{2\bar{\mu}\gamma_{0}t}\right)\nonumber \\
 &  & \times e^{-2[\gamma_{0}t-(\Delta\mu\gamma_{0}t)^{2}]},\label{eq:cr-cond}\\
z_{i}(t) & = & |\Phi_{+}\Phi_{-}|\sin\left(2\bar{\nu}\gamma_{0}t+\varphi\right)\nonumber \\
 &  & \times e^{-2[\gamma_{0}t+(\Delta\nu\gamma_{0}t)^{2}]},\label{eq:ci-cond}\end{eqnarray}
 where \[
\varphi=\arg(\Phi_{+}\Phi_{-}^{\ast})\]
 and the mean and variance are defined as \begin{eqnarray*}
\bar{\mu} & = & \int d^{3}rp_{{\rm rel}}(\vec{r})\mu(\vec{r}),\\
\Delta\mu^{2} & = & \int d^{3}rp_{{\rm rel}}(\vec{r})\mu^{2}(\vec{r})-\bar{\mu}^{2},\end{eqnarray*}
 and correspondingly for $\bar{\nu}$ and $\Delta\nu$.

\subsection{Generic conditions for finite-time disentanglement}

Finite-time disentanglement at a time $t_{d}$ requires that \[
z_{r}(t_{d})=z_{i}(t_{d})=0.\]
Let us first consider the case where $\bar{\mu}\neq0$. Then the real
part $z_{r}$ vanishes at the single time \begin{equation}
\gamma_{0}t_{d}=\ln(|\Phi_{+}|/|\Phi_{-}|)/2\bar{\mu}.\label{eq:td-cond}\end{equation}
Thus, given $\bar{\mu}\gtrless0$ we require an initial electronic
preparation with $|\Phi_{+}|\gtrless|\Phi_{-}|$ and $|\Phi_{\pm}|\neq0$
to obtain a positive and finite time $t_{{\rm d}}$. To accomplish
also a vanishing imaginary part, $z_{i}(t_{d})=0$, the phase $\varphi$
of the initial electronic preparation has to be adopted to compensate
for the accumulated phase due to the dipole-dipole interaction in
the sine of Eq. (\ref{eq:ci-cond}). This condition requires the phase
to be\[
\varphi=n\pi-\frac{\bar{\nu}}{\bar{\mu}}\ln(|\Phi_{+}|/|\Phi_{-}|),\qquad n\in\mathbb{Z}.\]
Therefore, there can always be found an initial electronic preparation
for which finite-time disentanglement occurs at exactly one time given
by Eq. (\ref{eq:td-cond}). At this time the concurrence vanishes
but immediately revives, similar to Ref. \cite{orszag}.

However, if $\bar{\mu}=0$, a vanishing real part $z_{r}$ can only
be obtained for \[
|\Phi_{+}|=|\Phi_{-}|,\]
corresponding to the initial electronic state\begin{equation}
|\Psi\rangle_{{\rm el}}=\cos({\textstyle \frac{\varphi}{2}})|{\textstyle -\frac{1}{2}},{\textstyle \frac{1}{2}}\rangle_{{\rm el}}+i\sin({\textstyle \frac{\varphi}{2}})|{\textstyle \frac{1}{2}},{\textstyle -\frac{1}{2}}\rangle_{{\rm el}}.\label{eq:state-mu0}\end{equation}
As now $z_{r}=0$ for all times, if $\bar{\nu}\neq0$, the time of
disentanglement $t_{{\rm d}}$ will be determined by the condition
$z_{i}(t_{{\rm d}})=0$, which results in\begin{equation}
\gamma_{0}t_{{\rm d}}=(n\pi-\varphi)/2\bar{\nu},\qquad n\in\mathbb{Z}.\label{eq:series-ftd}\end{equation}
Thus, for $\bar{\mu}=0$ and $\bar{\nu}\neq0$, within the range of
validity $\gamma_{0}t_{{\rm d}}\ll1$, a series of equidistant finite
times of disentanglement exists given an initial state of the form
(\ref{eq:state-mu0}). \textcolor{red}{}However, the periodic revivals
of the concurrence are due to the unitary and thus coherent dipole-dipole
interaction between the atoms, cf. Eq. (\ref{eq:ci-cond}). Thus,
different to the usual FTD, no decoherence can be attributed to them. 

On the other hand, if $\bar{\mu}=0$ and also $\bar{\nu}=0$ the condition
$z_{i}(t_{{\rm d}})=0$ will lead to $\varphi=n\pi$ ($n\in\mathbb{Z}$),
which corresponds for the required initial state (\ref{eq:state-mu0})
to a perfect separability of the two atoms throughout their entire
evolution in time. Therefore, we conclude that for $\bar{\mu}=\bar{\nu}=0$
finite-time disentanglement does \emph{not} exist.

\subsection{Dependence on distance and localization of the atoms}

Let us now apply these conditions considering the actual form of the
functions $\mu$ and $\nu$, and their averaging with the normalized
Gaussian $p_{{\rm rel}}$ with mean distance $r_{0}$ and rms spread
$\Delta r_{0}$:

\paragraph*{Far field ---}

In the far field, $r_{0}\gg\lambda_{0}$, both $\mu$ and $\nu$ asymptotically
decay to zero, so that as a consequence their mean values vanish,
$\bar{\mu},\bar{\nu}\to0$, quite independently of the spread $\Delta r_{0}$.
Therefore, finite-time disentanglement does \emph{not} exist for $r_{0}\gg\lambda_{0}$,
which is in agreement with our intuition: For distances larger than
the correlation length of the electromagnetic vacuum the atoms interact
with two statistically independent reservoirs, in which case finite-time
disentanglement does not exist if maximally one atom is initially
excited.

\begin{figure}
\begin{centering}\includegraphics[width=0.45\textwidth]{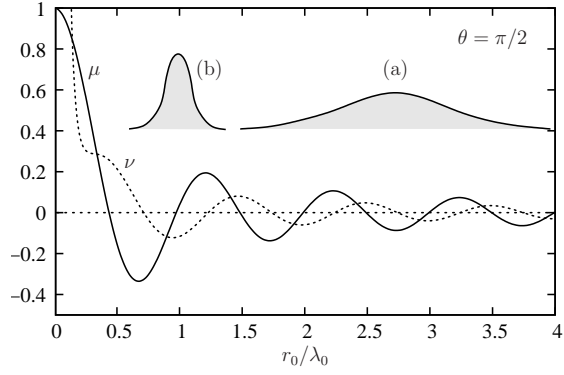}\par\end{centering}

\caption{Dependence of $\mu$ and $\nu$ on the inter atomic distance $r$.
The shaded Gaussians indicate the averaging over the inter atomic
distance with $\Delta r_{0}\gg\lambda_{0}$ (a) and $\Delta r_{0}<\lambda_{0}$
(b). For angles $\theta\neq\pi/2$ a more rapid decay would be observed
and at $r\to0$ the potential energy $\propto\nu$ would be attractive
instead of repulsive.}

\label{fig:mu-nu} 
\end{figure}

\paragraph*{Near field ---}

For a mean distance of the order of the wavelength, $r_{0}\sim\lambda_{0}$,
the behavior of $\mu$ and $\nu$ is dominated by oscillations with
period $\lambda_{0}$, see Fig. \ref{fig:mu-nu}. In this case, for
decreasing mean distance, the atoms are supposed to start to interact
with a common reservoir. There are now two possible scenarios, where
$\bar{\mu}$ may vanish:

(a) The spread is $\Delta r_{0}>\lambda_{0}$ but still $\Delta r_{0}\ll r_{0}$
so that the averaging is over at least one oscillation of $\mu$ and
$\nu$, leaving vanishing mean values $\bar{\mu}\approx\bar{\nu}\approx0$,
see Fig. \ref{fig:mu-nu} with inset (a). Also in this case finite-time
disentanglement does \emph{not} exist. This case corresponds to a
distance between atoms, that is not well localized in space with respect
to the wavelength $\lambda_{0}$, so that distance-dependent reservoir-mediated
effects are washed out. This behavior has not been seen in previous
work \cite{tanas,orszag} as it uniquely arises from the quantumness
of atomic positions.

(b) The spread is $\Delta r_{0}<\lambda_{0}$ and $r_{0}$ is centered
near to a node of $\mu$, leading to $\bar{\mu}=0$, see Fig. \ref{fig:mu-nu}
with inset (b). As the nodes of $\nu$ are approximately shifted with
respect to those of $\mu$ by $\lambda_{0}/4$, their mean will not
vanish in this case: $\bar{\nu}\neq0$. Thus, a series of equidistant
finite disentanglement times according to Eq. (\ref{eq:series-ftd})
will be observed. Such a repeated disentanglement occurs also in case
of two initial excitations, cf. Refs \cite{tanas,eberly3}. 

In all other cases of a near-field reservoir-mediated interaction,
a single finite disentanglement time according to Eq. (\ref{eq:td-cond})
exists. Thus, a high sensibility on the positioning and localization
of the atoms in the near-field is revealed. Only for distances $r_{0}\sim\lambda_{0}$
finite-time disentanglement can exist, because only in the near field
the atoms are located in a {}``common'' reservoir. However, only
well localized atoms with $\Delta r_{0}<\lambda_{0}$ can show this
peculiar behavior, otherwise the distance dependent coupling is washed
out. Furthermore, given well localized atoms in the near field, the
number of finite disentanglement times for a given initial state depends
on the precise distance between the atoms: If $r_{0}$ is at a node
of $\mu$ and if the initial state is of the form (\ref{eq:state-mu0}),
a series of equidistant finite disentanglement times exists. For other
distances only a single finite disentanglement time exists.

\section{Summary and conclusions\label{sec:Summary-and-conclusions}}

Among the various discussions of finite-time disentanglement for two-atom
systems, the work of Ficek and Tana\'{s} \cite{tanas} is closest
to our approach. However, there, an initial state including \emph{two
excitations}, i.e. both atoms being initially excited, was studied.
Moreover, the inter atomic distance was treated classically, thereby
discarding quantum dispersion and photon recoil. Our results can reproduce
this approximation by taking the limit $\Delta r_{0}\to0$, which,
however, is incompatible with the requirement of distinguishability,
see condition (\ref{eq:dr-limits}).

A further, more drastic approximation is that of disregarding the
relative position entirely and specifying either common or statistically
independent reservoirs for the two atoms. Such approximations can
be obtained from our results as limiting cases, further discarding
the dipole-dipole interaction ($\nu\to0$): In the limit $r_{0}\to0$
and thus $\gamma(\vec{r})\to\gamma_{0}$ a common reservoir is reproduced,
whereas for $r_{0}\to\infty$ and thus $\gamma(\vec{r})\to0$ two
statistically independent reservoirs emerge. The former case reveals
finite-time disentanglement for an initial single excitation \cite{orszag}.
It is, however, inconsistent in discarding the dipole-dipole interaction
at small distances. The latter case, on the other hand, does not show
finite-time disentanglement for a single initial excitation.

In conclusion, our results offer a consistent treatment of finite-time
disentanglement of two atoms with initial states containing no more
than a single excitation. Only in the near field, $r_{0}\sim\lambda_{0}$,
and for sufficiently well localized atoms a finite-time disentanglement
can be observed. The permitted range of rms spreads is $\Delta r_{{\rm min}}\ll\Delta r_{0}<\lambda_{0}$,
where the lower limit ensures the distinguishability of the atoms
during the observation time and the upper limit allows for resolving
the distance-dependent reservoir-mediated coupling between atoms.
If the distance $r_{0}$ is at a node of $\mu$, a series of equidistant
finite disentanglement times is observed for a particular type of
initial electronic state, whereas for other distances only a single
finite time of disentanglement can exist.

\begin{acknowledgments}
The authors thank J.H. Eberly, S. Maniscalco, and J. Piilo for discussions.
Financial support is acknowledged from FONDECYT grants nos. 3085030
(F.L.), 1051072 (S.W.), 1051062 (M.O.), and CONICYT doctoral fellowship
(M.H.). 
\end{acknowledgments}

\end{document}